\documentclass{ECOC-author}
\usepackage{etex}
\usepackage{amsmath,amssymb}
\usepackage{fst}
\usepackage{ctable}
\usepackage{makecell}

\newcommand{\w}{w}

\usepackage{soul}

\begin{document}

\title{Coding for Optical Communications -- Can We Approach the Shannon Limit With Low Complexity?}

\author{(Invited paper)\\\vspace{2mm}Alexandre Graell i Amat\ad{1*}, Gianluigi Liva\ad{2}, and  Fabian Steiner\ad{3}}
	\address{\add{1}{Department of Electrical Engineering, Chalmers University of Technology, Gothenburg, Sweden}
		\add{2}{Institute of Communications and Navigation of the German Aerospace Center (DLR), Munich, Germany}
        \add{3}{Institute for Communications Engineering, Technical University of Munich, Munich, Germany}
		\email{alexandre.graell@chalmers.se}}


\begin{abstract}
\vspace{-2mm}
Approaching capacity with low complexity is a very challenging task. In this paper, we review and compare three promising coding solutions to achieve that, which are suitable for future very high-throughput, low-complexity optical communications.
\end{abstract}

\maketitle

\section{Introduction}

Since the outset of forward error correction (FEC) for fiber-optic communications, research has intensively pursued the quest for approaching the theoretical limits. The prevailing choice in early fiber-optic systems was hard decision (HD) FEC  due to the lack of analog-to-digital converters (ADCs) and the requirement of very simple high-speed receivers. Coding schemes such as product codes (PCs) \cite{Eli54} and staircase codes \cite{Smi12,Zha14}, with low-complexity HD decoding based on bounded distance decoding (BDD) of the component codes, already provide significant coding gains at the required very low error rates. However, they still perform relatively far away from  the channel capacity. 

The advent of coherent transmission schemes and high resolution ADCs enabled the use of soft decision (SD) FEC. This has led to the progressive adoption of powerful coding schemes such as \ac{LDPC} codes and more recently spatially coupled \ac{LDPC} (SC-LDPC) codes \cite{Sug13}, which under SD decoding yield performance very close to the theoretical limits. 
However, while the complexity of SD-FEC  may be tolerable for long-haul communications and its excellent performance leaves little room for improvement,
several applications, such as metro networks and data center interconnects,  require very high throughputs (in the order of several hundreds of Gbps or even Tbps) and low power consumption. Scaling SD-FEC schemes to such high throughputs and low power consumption is a very challenging task and hence these schemes  are not suited for these applications
. Thus, a fundamental question is: Can we approach the Shannon limit yet with low complexity, i.e., while still achieving the high-throughput and low power consumption of the HD-FEC solutions? Answering this question requires envisaging new coding schemes and decoding methods and gives a unique opportunity for groundbreaking contributions.  



In this paper, we review and compare three important approaches that are currently being considered to address the fundamental question above and we highlight some of the open research questions that need to be addressed.
An important observation is that a main limiting factor to achieve very high throughputs with SD-FEC is the high internal data flow in the decoder due to the exchange of soft messages, rather than the resolution of the ADC. This observation is at the basis of the binary message passing (BMP) decoding algorithm introduced in \cite{Lec12} for \ac{LDPC} codes. The key idea of BMP is to exploit the channel soft information  while only exchanging binary messages among the component codes decoders during iterations. This idea was extended in \cite{Yac19,Ste19} to two-bit messages, giving rise to ternary message passing (TMP) and quaternary message passing (QMP) algorithms. In general, an interesting approach is to consider coarsely quantized \ac{LDPC} decoders, where the exchanged messages are limited to a small number of bits to keep the decoder data flow low. Another line of research  is to consider coding schemes based on HD decoding and assist the decoding with some level of soft information to improve its performance, while keeping the decoder data flow identical or close to that of conventional HD decoding based on BDD of the component codes. Several soft-aided decoding algorithms have recently been proposed for product-like codes, see, e.g., \cite{Hag18,She18,She19,She19t,Lei19}. A third alternative approach, explored in \cite{Zha17,Bar18}, is to consider a hybrid HD-SD FEC scheme based on an inner SD-FEC code and an outer HD-FEC code. The key idea is that the inner code, designed such that its decoder fulfils a given complexity constraint, is used for error reduction at the input of the outer HD-FEC (a staircase code) decoder, which then takes care of lowering the error rate to the desired target.

\section{Hybrid SD-HD Schemes}

The main idea of the hybrid scheme in \cite{Zha17,Bar18} is to combine the close-to-capacity performance of SD FEC with sparse-graph codes with the low complexity of HD FEC. In particular, a key observation is that the high complexity of state-of-the-art SD FEC solutions stems from performing close to capacity at very low error probability.  However, if only moderate error probabilities are sought for, low-complexity SD FEC codes can be designed to achieve these. The hybrid scheme in \cite{Zha17,Bar18} builds on this observation by concatenating a relatively weak, low complexity inner SD code with an outer HD staircase code. The main task of the inner code (an \ac{LDPC} code in \cite{Bar18}) is to reduce the bit error probability below the threshold of the outer code, which corrects the majority of the errors. As the inner code only needs to achieve a moderate probability of error (in the range $10^{-2}$--
$10^{-3}$), it can be of low complexity. In  \cite{Zha17,Bar18} the inner code is designed such that a complexity score based on the number of edges in its Tanner graph and the number of decoding iterations is minimized. This design approach results in a Pareto frontier that characterizes the trade-off between coding gain and complexity. The scheme proposed in \cite{Zha17,Bar18} achieves similar coding gains to those of existing soft-decision FEC schemes with a significant reduction in complexity.

\section{Coarsely Quantized LDPC Decoders}

Iterative decoding of \ac{LDPC} codes entails an iterative message exchange between variable nodes (VNs) and check nodes (CNs) in the code graph. The amount of information transferred in each iteration is proportional to the product $n \bar{d}_\mathtt{v} q$, where $n$ is the block length (i.e., the number of VNs in the code graph), $\bar{d}_\mathtt{v}$ is the average VN degree, and $q$ is the number of bits used to represent each message. It was recognized in \cite{Smi12} that this quantity represents the actual limiting factor in the implementation of very high throughput decoders. An obvious consequence of this observation is that, to develop efficient LDPC decoder implementations targeting speeds of several hundred Gbps, the main decoder design parameter to play with is the message quantization.

Current high speed implementations adopt $4$ or $5$ bits per message \cite{Smi12}. When this number of quantization bits is used, ad-hoc CN and VN update rules may be developed to limit the loss with respect to unquantized belief propagation (BP) decoding. Examples of approaches addressing the decoder design are given, for instance, by the application of the information bottleneck method \cite{Bau18} or by the definition of global cost functions such as the iterative decoding threshold \cite{Ste18} or the error rate at which the error floor emerges \cite{Pla11}. 

A possibility to further improve the decoding speed relies in a further reduction of the number of bits used to represent each message. In the extreme case, each message may be represented by one bit only, giving rise to the family of BMP algorithms. The simplest examples of LDPC codes decoding algorithms relying on one-bit messages were introduced by Gallager \cite{Gal63} (the so-called Gallager A and Gallager B algorithms). However, both algorithms introduced in \cite{Gal63} were designed to operate with hard decisions from the channel. In \cite{Lec12} a key modification was introduced in BMP decoding, enabling the use of the soft information available at the channel output. The algorithm proposed in \cite{Lec12}  works as follows. The messages exchanged between VNs and CNs belong to the binary alphabet $\mathcal{M}=\{-1,+1\}$.
Denote by $m_{\mathtt{v} \rightarrow \mathtt{c}}^{(\ell)}$ the message sent by  VN $\mathtt{v}$ to  CN $\mathtt{c}$ and by $m_{\mathtt{c} \rightarrow \mathtt{v}}^{(\ell)}$ the message sent by $\mathtt{c}$ to $\mathtt{v}$ during the $\ell$-th iteration. Denote furthermore the channel log-likelihood ratio (LLR) at the input of $\mathtt{v}$ as $L$. The message from $\mathtt{v}$ to $\mathtt{c}$ is obtained by combining the channel soft-information $L$ with a weighted version of all other incoming CN messages. Finally, a hard decision is applied to turn the result into a binary message, i.e., we have
\[
m_{\mathtt{v} \rightarrow \mathtt{c}}^{(\ell)}=f\left(L+\sum_{\mathtt{c}'\neq\mathtt{c}}w^{(\ell-1)}m_{\mathtt{c}' \rightarrow \mathtt{v}}^{(\ell-1)}\right)
\]
where $f(x)=+1$ if $x>0$, and $f(x)=-1$ otherwise. The weighting factors $w^{(\ell)}$ are real valued and depend on the current iteration number. They can be obtained from the \ac{DE} analysis as proposed in \cite{Lec12}. The update rule at the CNs involves a multiplication of the incoming messages. The algorithm enables substantial gains with respect to the algorithms originally proposed in \cite{Gal63}, limiting the performance with respect to unquantized SD decoding to a few tenths of a dB (especially at high code rates). In \cite{Yac19,Ste19} we introduced extensions of BMP toward messages belonging to ternary and quaternary sets (i.e., $2$-bit message passing algorithms \cite{Sas09}). The two algorithms, TMP and QMP, enable to half the gap in coding gain between BMP and unquantized BP.

\section{Soft-Aided Decoding of Product-Like Codes}

An alternative to the approaches in the previous sections is to consider coding schemes based on HD decoding and enhance the basic HD decoder while limiting the increase in complexity. In \cite{Hag18}, a decoding algorithm that exploits conflicts between component codes in order to assess their reliabilities, even when no channel reliability information is available, is proposed. The main idea of the algorithm, dubbed anchor decoding (AD), is to introduce status information for each component code and designate certain ``reliable'' component codes as anchors.

\begin{figure}[!t] \centering 
 	\includegraphics[width=\columnwidth]{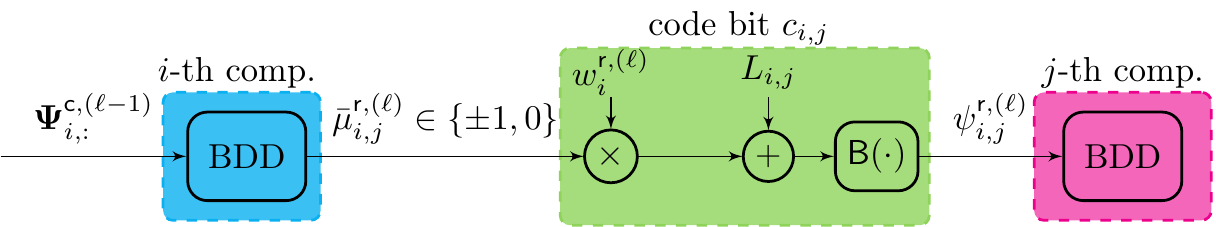}  
 	\vspace{-4.5ex}
 	\caption{\footnotesize Block diagram of iBDD-SR.}  
 	\label{SysPCCST} 
	\vspace{-2ex}
 \end{figure}

Extending the BMP algorithm in \cite{Lec12} to product-like codes, in \cite{She18,She19t} we proposed a new decoding algorithm, called iBDD with scaled reliability (iBDD), which exploits some level of soft information but keeps the messages exchanged between component decoders binary. The main principle of the iBDD-SR is to make a hard decision on a weighted sum of the BDD output with the channel LLR, where the BDD decoder output reliability is conveyed by a scaling factor applied to the BDD outbound messages. 
iBDD-SR is illustrated in Fig.~\ref{SysPCCST}.
Consider the decoding of the $i$-th row code at iteration $\ell$ and assume without loss of generality that zeroes are represented by $-1$ and ones by $+1$. First, BDD is performed based on the hard decisions after the decoding of the column codes at iteration $\ell-1$, collected in the matrix  $\boldsymbol{\Psi}^{\mathsf{c},{(\ell-1)}}_{i,:}$. 
To exploit soft information, the output of the BDD stage takes values on a ternary alphabet $\{\pm 1,0\}$ where $0$ corresponds to a decoding failure. The reliability information on code bit $c_{i,j}$ (code bit in the $i$-th row and $j$-th column of the code array) is then formed according to
$\mu_{i,j}^{\mathsf r, (\ell)}=\w_i^{\mathsf r, (\ell)} \cdot \bar{\mu}_{i,j}^{\mathsf r,(\ell)} + L_{i,j}$, where $L_{i,j}$ is the channel LLR and $\bar{\mu}_{i,j}^{\mathsf r,(\ell)}$ is the output of the BDD for code bit $c_{i,j}$ . $w_\ell>0$ is a scaling factor that can be optimized via density evolution \cite{She19t}. Finally, a hard decision is made on $\mu_{i,j}^{\mathsf r, (\ell)}$, and the hard decision $\psi_{i,j}^{\mathsf{r},(\ell)}=
f(\mu_{i,j}^{\mathsf r, (\ell)})$ is passed to the $j$-th column code.

While iBDD-SR exploits soft information and thus is a SD decoding algorithm in nature, the component decoders solely exchange hard decision, hence the algorithm yields the same data flow of that of iBDD-SR and a negligible increase in complexity. For a PC with $(255,231,3)$ BCH component codes, iBDD-SR was implemented in \cite{Fou19} with 28nm process technology, achieving $1$ Tbs with $0.2$ dB gain compared to conventional HD decoding. Furthermore, PCs with iBDD-SR achieve similar coding gains than staircase codes with an area and energy dissipation less than half.

Other algorithms that exploit some level of soft information have been proposed, see, e.g., \cite{She19,Lei19}. However, these algorithm require a sorting of the least reliable bits after each row and column decoding, and their decoding complexity needs to be further investigated.

\section{Numerical Results}

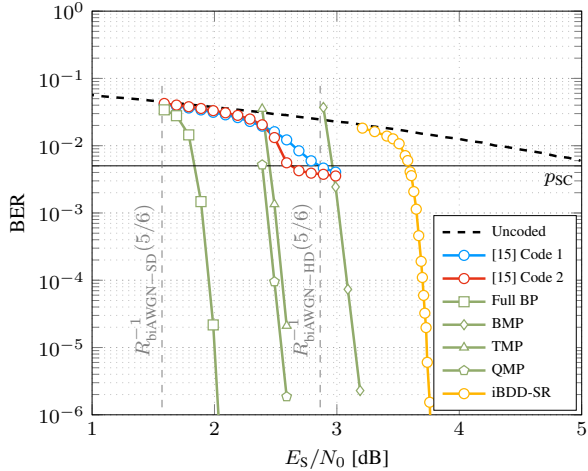
\begin{figure}[!t]
\footnotesize
\centering
\begin{tikzpicture}[scale=0.94]
\begin{axis}[
xlabel={$E_\tS/N_0$ [dB]},
ylabel={BER},
ymode=log,
grid style={gray,opacity=0.5,dotted},
grid=both,
legend cell align={left},
legend pos=south east,
legend style={at={(0.99,0.01)},anchor=south east},
xmin=1,
xmax=5,
ymin=1e-6,
ymax=1,
]

\addplot[name path global=uncoded,line width=1,dashed,x filter/.code={\pgfmathparse{\pgfmathresult-3.0103}\pgfmathresult}] table[x=snr,y=ber] {data/uncoded_ber_biawgn.txt};
\addlegendentry{\tiny Uncoded};

\addplot[name path global=kschi1,line width=1,TUMBeamerBlue,mark=*, fill=white,mark options={solid, line width = 0.5pt, fill=white},x filter/.code={\pgfmathparse{\pgfmathresult-3.0103}\pgfmathresult}] table[x=snr,y=ber] {data/results_kschi1.txt};
\addlegendentry{\tiny \cite{Bar18} Code 1};
\addplot[name path global=kschi2,line width=1,TUMBeamerRed,mark=*,mark options={solid, line width = 0.5pt, fill=white},x filter/.code={\pgfmathparse{\pgfmathresult-3.0103}\pgfmathresult}] table[x=snr,y=ber] {data/results_kschi2.txt};
\addlegendentry{\tiny \cite{Bar18} Code 2};

\addplot[name path global=full,line width=1,TUMBeamerGreen,mark=square*,mark options={solid, line width = 0.5pt, fill=white},x filter/.code={\pgfmathparse{\pgfmathresult-3.0103}\pgfmathresult}] table[x=snr,y=ber] {data/results_sc-dv=4_dc_24_L=50_Q=320_full.txt};
\addlegendentry{\tiny Full BP};

\addplot[name path global=full,line width=1,TUMBeamerGreen,mark=diamond*,mark options={solid, line width = 0.5pt, fill=white},x filter/.code={\pgfmathparse{\pgfmathresult-3.0103}\pgfmathresult}] table[x=snr,y=ber] {data/results_sc-dv=4_dc_24_L=50_Q=320_bmp.txt};
\addlegendentry{\tiny BMP};

\addplot[name path global=full,line width=1,TUMBeamerGreen,mark=triangle*,mark options={solid, line width = 0.5pt, fill=white},x filter/.code={\pgfmathparse{\pgfmathresult-3.0103}\pgfmathresult}] table[x=snr,y=ber] {data/results_sc-dv=4_dc_24_L=50_Q=320_tmp.txt};
\addlegendentry{\tiny TMP};

\addplot[name path global=full,line width=1,TUMBeamerGreen,mark=pentagon*,mark options={solid, line width = 0.5pt, fill=white},x filter/.code={\pgfmathparse{\pgfmathresult-3.0103}\pgfmathresult}] table[x=snr,y=ber] {data/results_sc-dv=4_dc_24_L=50_Q=320_qmp.txt};
\addlegendentry{\tiny QMP};


\addplot[name path global=iBDD-SR,line width=1,TUMBeamerYellow,mark=*,mark options={solid, line width = 0.5pt, fill=white},x filter/.code={\pgfmathparse{\pgfmathresult-0.7915}\pgfmathresult}] table[x=ebno,y=ber] {data/results_pc_fixed.txt};
\addlegendentry{\tiny iBDD-SR};






\draw[] (axis cs:0,5.02e-3) -- (axis cs:8,5.02e-3) node[below,midway,xshift=1.4cm] {$p_{\tSC}$};




\draw[gray, dashed] (axis cs:1.5713,1e-7) -- (axis cs:1.5713,1e-1) node[sloped,midway,yshift=7pt] {$R_{\tbiawgn-\tSD}^{-1}(5/6)$};
\draw[gray, dashed] (axis cs:2.8633,1e-7) -- (axis cs:2.8633,1e-1) node[sloped,midway,yshift=7pt] {$R_{\tbiawgn-\tHD}^{-1}(5/6)$};

\end{axis}
\end{tikzpicture}
\vspace{-3.5ex}
\caption{\footnotesize Finite length performance of different \ac{FEC} and decoding solutions. The \ac{FEC} overhead is $\approx 20\%$ and the codelength is about \num{96000} bits.}
\label{fig:coded_results}
\vspace{-2.5ex}
\end{figure}

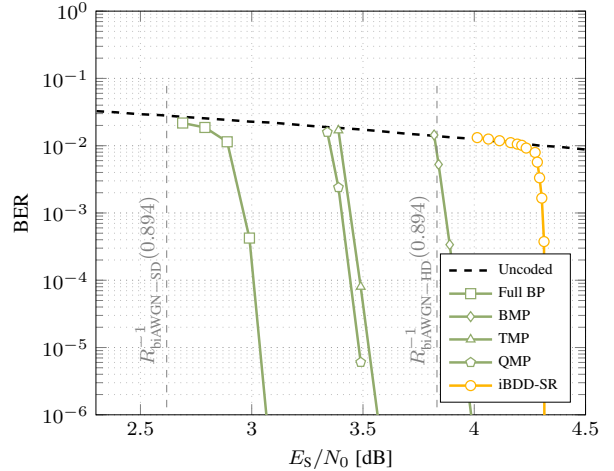
\begin{figure}[!t]
\footnotesize
\centering
\begin{tikzpicture}[scale=0.94]
\begin{axis}[
xlabel={$E_\tS/N_0$ [dB]},
ylabel={BER},
ymode=log,
grid style={gray,opacity=0.5,dotted},
grid=both,
legend cell align={left},
legend pos=south east,
xmin=2.3,
xmax=4.5,
ymin=1e-6,
ymax=1,
]

\addplot[name path global=uncoded,line width=1,dashed,x filter/.code={\pgfmathparse{\pgfmathresult-3.0103}\pgfmathresult}] table[x=snr,y=ber] {data/uncoded_ber_biawgn.txt};
\addlegendentry{\tiny Uncoded};
 
\addplot[name path global=full,line width=1,TUMBeamerGreen,mark=square*,mark options={solid, line width = 0.5pt, fill=white},x filter/.code={\pgfmathparse{\pgfmathresult-3.0103}\pgfmathresult}] table[x=snr,y=ber] {data/results_sc_dv=4_dc=40_L=50_Q=520_full.txt};
\addlegendentry{\tiny Full BP};

\addplot[name path global=full,line width=1,TUMBeamerGreen,mark=diamond*,mark options={solid, line width = 0.5pt, fill=white},x filter/.code={\pgfmathparse{\pgfmathresult-3.0103}\pgfmathresult}] table[x=snr,y=ber] {data/results_sc_dv=4_dc=40_L=50_Q=520_bmp.txt};
\addlegendentry{\tiny BMP};

\addplot[name path global=full,line width=1,TUMBeamerGreen,mark=triangle*,mark options={solid, line width = 0.5pt, fill=white},x filter/.code={\pgfmathparse{\pgfmathresult-3.0103}\pgfmathresult}] table[x=snr,y=ber] {data/results_sc_dv=4_dc=40_L=50_Q=520_tmp.txt};
\addlegendentry{\tiny TMP};

\addplot[name path global=full,line width=1,TUMBeamerGreen,mark=pentagon*,mark options={solid, line width = 0.5pt, fill=white},x filter/.code={\pgfmathparse{\pgfmathresult-3.0103}\pgfmathresult}] table[x=snr,y=ber] {data/results_sc_dv=4_dc=40_L=50_Q=520_qmp.txt};
\addlegendentry{\tiny QMP};

\addplot[name path global=iBDD-SR,line width=1,TUMBeamerYellow,mark=*,mark options={solid, line width = 0.5pt, fill=white},x filter/.code={\pgfmathparse{\pgfmathresult-0.4866}\pgfmathresult}] table[x=snr,y=ber] {data/results_sc.txt};
\addlegendentry{\tiny iBDD-SR};

\draw[gray, dashed] (axis cs:2.6180,1e-7) -- (axis cs:2.6180,1e-1) node[sloped,midway,yshift=7pt] {$R_{\tbiawgn-\tSD}^{-1}(0.894)$};
\draw[gray, dashed] (axis cs:3.8324,1e-7) -- (axis cs:3.8324,1e-1) node[sloped,midway,yshift=7pt] {$R_{\tbiawgn-\tHD}^{-1}(0.894)$};

\end{axis}
\end{tikzpicture}
\vspace{-3.5ex}
\caption{\footnotesize Comparison of finite length performance of different \ac{FEC} and decoding solutions. The \ac{FEC} overhead is $\approx 11.86\%$.}
\label{fig:coded_results1}
\vspace{-2.5ex}
\end{figure}

In the following, we compare the performance of the different \ac{FEC} and decoding architectures for transmission over a binary-input AWGN channel
such that the channel output is $Y = X + N$.
The channel input $X$ is uniformly distributed on $\{-1,+1\}$ and the noise $N$ is a Gaussian random variable with zero mean and variance $\sigma^2$, i.e., $N\sim \cN(0,\sigma^2)$. We have $E_\tS/N_0 = 1/(2\sigma^2)$.
In Fig.~\ref{fig:coded_results}, we consider a setting with a target \ac{OH} of 20\%, i.e., a code rate of $5/6$. The codes have a blocklength of around \num{96000} bits. The respective Shannon limits for SD and HD decoding are given
by dashed vertical lines. The SC-LDPC code is regular with degree four VNs. It is terminated after 50 spatial positions and has an effective code rate of \num{0.8233}. We decode the code by sum-product BP, BMP, TMP
and QMP. At a BER of \num{e-6}, we see a gain of \SI{0.5}{dB} and \SI{0.6}{dB} for TMP and QMP over BMP. The gap to unquantized BP is \SI{0.7}{dB} for QMP. We also show the performance of two proposed SD inner code designs from \cite[Sec.~4.2]{Bar18} with different complexity scores. The crossing of these curves with the horizontal line at a BER of \num{5.02e-3} indicates the $E_\tS/N_0$ to further drive the BER down to \num{e-15} after the HD outer staircase code. We observe that code design~1 looses about \SI{0.3}{dB} in power efficiency compared to code design~2, however it achieves the HD FEC threshold with lower complexity. We also show the performance of the iBDD-SR scheme for a PC with (310, 283, 3) BCH component codes. In Fig.~\ref{fig:coded_results1}, we depict a scenario for an OH of 11.86\% (code rate \num{0.894}). We compare a terminated SC-LDPC code  with 50 spatial positions (resulting blocklength is \SI{260000}{bits}), decoded with Full BP, BMP, TMP and QMP, with a staircase code based on (510, 483, 3) BCH component codes. At a BER of \num{e-6}, BMP and QMP have a gap of \SI{0.9}{dB} and \SI{0.5}{dB} to unquantized BP decoding, i.e., the observed gaps become even smaller for higher code rates. 

\section{Conclusion}

Designing FEC coding/decoding schemes  that are able to approach the Shannon limit with low complexity is important to enable future very high-throughput, low power fiber-optic communication systems. It is perhaps too audacious to claim  that this goal has already been achieved. In this paper, however, we discussed three very promising research lines that represent important steps toward this goal. While from an algorithmic complexity perspective the discussed FEC solutions are very appealing, a thorough complexity evaluation, in particular of the attainable throughputs and power consumption is still required. This cannot be based solely on algorithmic considerations, but also requires considering memory access and wiring issues. Therefore, a joint effort of code/decoding design and implementation is required.
Overall, coding for high-throughput applications, such as fiber-optic communications, is a very timely, exciting research problem.

\section*{Acknowledgment}
\vspace{-1ex}
	\scriptsize{ The work of A.~Graell i Amat was partially supported by the Knut and Alice Wallenberg Foundation.}
	
\newpage

\section*{References}
	\normalsize

\end{document}